\documentclass[useAMS,usenatbib]{mn2e}
\usepackage{txfonts}
\usepackage[latin1]{inputenc}
\usepackage[T1]{fontenc}
\usepackage{ngerman}
\usepackage{mathptmx}
\usepackage[scaled]{helvet}
\usepackage{courier}
\usepackage{aecompl} 
\usepackage{longtable}  
\usepackage{rotating}
\usepackage{natbib}
\usepackage{graphicx}
\usepackage{graphics}
\usepackage{psfrag}
\usepackage{amssymb}
\usepackage{multicol}
\usepackage{multirow}
\usepackage{lscape}
\usepackage{color}
\usepackage{soul}
\bibpunct{(}{)}{;}{a}{}{,}
\def\Teff{\ensuremath{T_{\mathrm{eff}}}}
\def\logg{$\log{g}$}

\title[]{The first $\Delta a$ observations of three globular clusters} 

\author[E. Paunzen et al.]{E. Paunzen,$^{1}$\thanks{epaunzen@physics.muni.cz}
I.Kh.~Iliev,$^{2}$
O.I.~Pintado,$^{3}$
H.~Baum,$^{4}$
H.M.~Maitzen,$^{4}$
M.~Netopil,$^{1,4}$ \newauthor
A.~{\"O}nehag,$^{5}$
M.~Zejda,$^{1}$
and L.~Fraga$^{6,7}$
			   \\
$^{1}$Department of Theoretical Physics and Astrophysics, Masaryk University,
Kotl\'a\v{r}sk\'a 2, 611\,37 Brno, Czech Republic\\
$^{2}$Rozhen National Astronomical Observatory, Institute of Astronomy of the Bulgarian 
Academy of Sciences, PO Box 136, BG-4700 Smolyan, Bulgaria\\
$^{3}$Instituto Superior de Correlaci{\'o}n Geol{\'o}gica, Av. Per{\'o}n S/N, Yerba Buena, 4000 
Tucum{\'a}n, Argentina\\
$^{4}$Institut f{\"u}r Astrophysik der Universit{\"a}t Wien, T{\"u}rkenschanzstr. 17, A-1180 Wien, Austria\\
$^{5}$Department of Physics and Astronomy, Uppsala University, Box 516, S-751~20, Uppsala, Sweden\\
$^{6}$Southern Observatory for Astrophysical Research, Casilla 603, La Serena, Chile \\
$^{7}$Laborat{\'o}rio Nacional de Astrof{\'i}sica/MCTI, R. Estados Unidos, 154 Itajub{\'a}, MG, CEP: 37504-364, Brazil}
\begin{document}

\date{}

\pagerange{\pageref{firstpage}--\pageref{lastpage}} \pubyear{2014}

\maketitle

\label{firstpage}

\begin{abstract}
Globular clusters are main astrophysical laboratories to test and modify
evolutionary models. Thought to be rather homogeneous in their local elemental distribution
of members, results suggest a wide variety of chemical peculiarities. Besides different
main sequences, believed to be caused by different helium abundances, peculiarities of blue 
horizontal-branch stars and on the red giant branch were found. 
This whole zoo of peculiar objects has to be explained in the context of stellar formation 
and evolution. The tool of $\Delta a$ photometry 
is employed in order to detect peculiar stars in the whole spectral range.
This three filter narrow band system measures the flux distribution in the region from 
4900 to 5600\AA\, in order to find any peculiarities around 5200\AA. It is highly efficient to detect
classical chemically peculiar stars of the upper main sequence, Be/Ae, shell and metal-weak objects
in the Milky Way and Magellanic Clouds. We present $\Delta a$ photometry of 2266 stars from 109 individual 
frames for three globular clusters
(NGC~104, NGC~6205, and NGC~7099). A comparison with published abundances, 
for three horizontal-branch stars, only, yield an excellent agreement. 
According to the 3$\sigma$ detection limit of each globular cluster, about 3\% of the stars lie in 
abnormal regions in the diagnostic diagrams. 
The first observations of three widely different aggregates
give very promising results, which will serve as a solid basis for follow-up observations including
photometric as well as spectroscopic studies.
\end{abstract}


\section{Introduction}

Globular clusters are extensively used to place constraints on key ingredients of 
canonical stellar evolution models, such as the mixing length parameter of convective 
energy transport theory \citep{Ferr06}. They are also rather ``simple'' stellar systems
consisting of a distinct population, which is in dynamic equilibrium. Therefore, they are
used for extensive $N$-body simulation in order to understand the formation and evolution
of the Milky Way \citep{Sipp12}. 

It is now well known that several globular clusters have at least two main
sequences (MS), which are explained by a different helium content \citep{Piot07}. In addition,
different red giant (RGB) and sub-giant branches were also found \citep{Piot12}. 
A similar characteristic of the horizontal-branch (HB), namely at least two different populations, was detected by \citet{Grun98}.
There are different ``jumps'' in the blue HB (BHB) distribution in the $V$ versus $(u - y)$ colour-magnitude 
diagram, which could be used to select apparent peculiar objects \citep{Grun99}. Also, peculiar
HB extensions, like the blue-hook were found \citep{Brow01}. However,
the cause of these phenomena is still not clear, but it is probably connected to the 
complex star formation history of the individual clusters \citep{Valc12}.

\begin{table*}
\caption{The basic cluster parameters of the targets taken from \citet{Harr96}.}
\label{parameters_gcl}
\begin{center}
\begin{tabular}{lcccrccccc} 
\hline\hline 
NGC & Name & $\alpha$(2000) & $\delta$(2000) & \multicolumn{1}{c}{$l$} & $b$ & R$_{\sun}$ & $R_\mathrm{GC}$ & $E(B-V)$ & [Fe/H] \\
& & ($\degr$) & ($\degr$) & \multicolumn{1}{c}{($\degr$)} & ($\degr$) & (kpc) & (kpc) & (mag) & (dex) \\
\hline
104 & 47 Tuc & 00 24 05.67 & $-$72 04 52.6 & 305.89 & $-$44.89 & 4.5 & 7.4 & 0.04 & $-$0.72 \\ 
6205 & M13 & 16 41 41.24 & +36 27 35.5 & 59.01 & +40.91 & 7.1 & 8.4 & 0.02 & $-$1.53 \\
7099 & M30 & 21 40 22.12 & $-$23 10 47.5 & 27.18 &$-$46.84 & 8.1 & 7.1 & 0.03 & $-$2.27 \\
\hline 
\end{tabular}
\end{center}
\end{table*}

\citet{Grat12} gave an excellent overview of all types of chemical peculiarities for members
of globular clusters. Not only variations in light elements (Li, C, N, O, Na, Al, and Mg), but also of CN and
CH were detected. Possible correlations and relations of the individual abundances for different elements were
widely analysed in the context of stellar formation and evolution. In general terms, one has to distinguish,
as for Population I type objects, between intrinsic and photospheric peculiarities.

An intriguing phenomenon, which triggered our study, was first reported by 
\citet{Behr99} who found large deviations in element abundances from
the expected cluster metallicity for BHB stars in the globular 
cluster NGC~6205. For example, iron was found up to be a factor of 3 enhanced compared to the 
solar value, or about 100 times the mean cluster iron abundance.
Later investigations confirmed such elemental abundance anomalies for stars
of other globular clusters \citep{Pace06}. Those authors already speculated that elemental
diffusion in the stellar atmosphere will cause the phenomenon. \citet{Mich08} used stellar evolution 
models with self-consistent atomic diffusion to investigate the atmospheric effects in more detail.
Indeed, they were able to reproduce the observational results, confirming the role of atomic diffusion 
driven by radiative accelerations in HB stars.

Such atmospheric effects are well known and studied for the classical chemically peculiar (CP) stars of 
the upper MS \citep{Brai10}. Those similarities encouraged us to exploit the
capability to detect peculiar Population II stars with the tool of $\Delta a$
photometry. Due to the typical flux depression in CP stars at 5200\AA\,(full width at half-maximum, FWHM, of the 
corresponding filter is about 100\AA),
we are able to detect them in an economical and efficient way by comparing the flux at the 
centre (5200\AA, $g_{\rm 2}$) to the adjacent regions (5000\AA, $g_{\rm 1}$ and 5500\AA, 
$y$). It has been shown that virtually all peculiar stars with magnetic fields have significant
positive $\Delta a$ values of up to +100 mmag, whereas Be/Ae and metal weak stars exhibit significantly 
negative ones \citep{Paun05}. Furthermore, \citet{khan07} investigated the contribution of individual elements 
on the 5200\,\AA$ $ flux depression. They concluded that Fe is mainly responsible for
producing a positive $\Delta a$ index, but also Cr and Si are contributors at least for lower 
effective temperatures.

Inspecting the spectral region around 5200\AA, where the $g_{\rm 2}$ filter is centred, for RGB-,
G- and K-type MS stars \citep{Gray09}, we find strong Mg I (5167, 5173, and 5183\AA) and MgH features. 
Those features vary strongly with the evolutionary status, i.e. for dwarfs and giants.
\citet{John05} derived magnesium abundances for more than 100 RGB stars in each of the Galactic 
globular clusters NGC 5272 and NGC 6205. They found a wide spread of magnesium abundances within the individual
aggregates (about 0.7\,dex) and some significant outliers. So it might be possible to detect strongly deviating
elemental abundances of magnesium and luminosity effects with $\Delta a$ photometry.

A further motivation was its feasibility to detect CP stars 
in the Large Magellanic Cloud \citep{Paun06}, because its global environment is also very different
than in the Galactic disc. 

In this paper we present the first photometric $\Delta a$ observations of three, widely different, 
globular clusters NGC~104, NGC~6205, and NGC~7099. In this first observational pilot study,
we test, if the $\Delta a$ system is able to detect peculiar stars both at the HB- and RGB-, G- as well 
as K-type MS stars. We do not expect to find traces of the different star sequences due to the fact
that helium does not contribute in the 5200\AA\, region.

\begin{figure*}
\begin{center}
\includegraphics[width=160mm]{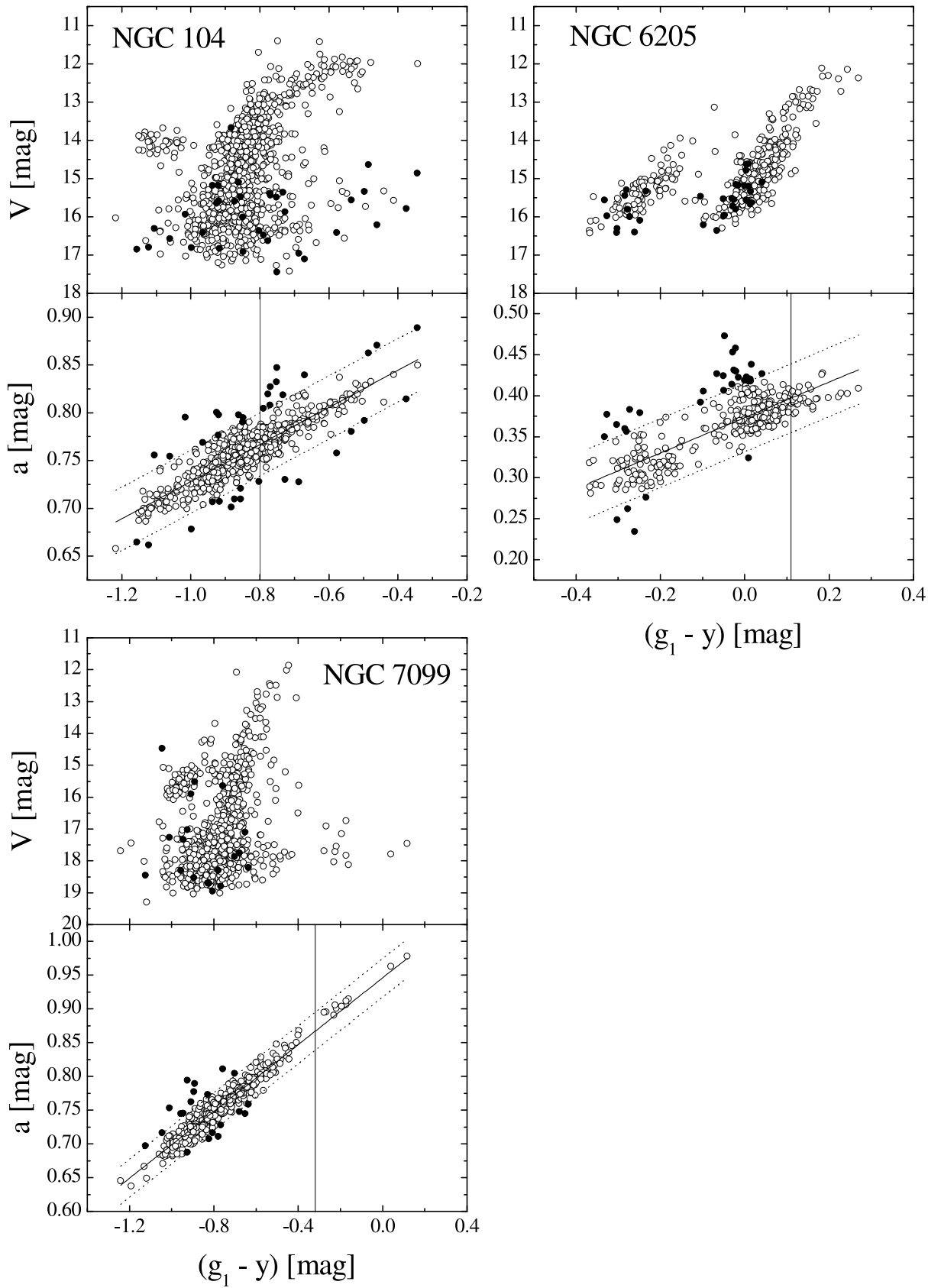}
\caption{The photometric diagrams of the three globular clusters. The dotted lines are the 
3$\sigma$ upper and lower limits of the $\Delta a$ detection sensitivity according to Table \ref{results_gcl}.
The vertical lines are the locus of $(B-V)_0$\,=\,1.0\,mag.}
\label{cmds}
\end{center}
\end{figure*}

\begin{table*}[t]
\begin{center}
\caption{Observing log for the programme clusters. The clusters were observed by 
I.Kh.~Iliev (II) and O.I.~Pintado (OP).}
\label{log}
\begin{tabular}{lccccccccc}
\hline\hline
Cluster & Site & Date & Obs. & \#$_{g_{\rm 1}}$ & \#$_{g_{\rm 2}}$ & \#$_{y}$ & $t_{g_{\rm 1}}$ & $t_{g_{\rm 2}}$ & $t_{y}$ \\
& & & & & & & (s) & (s) & (s) \\
\hline
NGC~104 & CASLEO & 08.2001 & OP & 6 & 10 & 10 & 5x60, 1x120 & 5x60, 5x120 & 5x120, 5x300 \\
NGC~6205 & BNAO & 07.2003 & II & 12 & 12 & 13 & 12x100 & 12x100 & 13x100 \\
NGC~7099 & CASLEO & 08.2001 & OP & 15 & 16 & 15 & 10x60, 6x120 & 10x60, 5x120 & 10x60, 5x180 \\
\hline
\end{tabular}
\end{center}
\end{table*}

\section{Target selection, observations and reduction}

For this first case study, we selected three globular clusters with
widely different overall metallicities, namely NGC~104, NGC~6205, 
and NGC~7099. Such a criterion guarantees to detect possible peculiar
stars for various local environments. Table \ref{parameters_gcl}
lists the basic cluster parameters of the targets, taken from 
\citet{Harr96}.

The observations of the three globular clusters were performed 
at two different sites: 
\begin{itemize}
\item 2\,m Ritchey-Chretien-Coude telescope (Bulgarian National Astronomical Observatory, BNAO, Rozhen),
direct imaging, SITe SI003AB 1024\,$\times$\,1024 pixel CCD,
5' field of view, 1 pixel\,=\,0.32$\arcsec$,
\item 2.15\,m telescope (El Complejo Astron{\'o}mico El Leoncito, CASLEO), direct imaging with focal reducer, 
TEK-1024 CCD, 9$\farcm$5 field of view, 1 pixel\,=\,0.813$\arcsec$.
\end{itemize}
The observing log with the number of frames in each filter and the integration times
is listed in Table \ref{log}. The typical seeing conditions were between 1 and 2
arc seconds.
The observations were performed with
two different filter sets, both having the following characteristics:
$g_1$ ($\lambda_{\rm c}$\,=\,5007\,\AA, FWHM\,=\,126\,\AA, $T_{\rm P}$\,=\,78\%), 
$g_2$ (5199, 95, 68) and $y$ (5466, 108, 70).

The CCD reductions were performed with standard IRAF v2.12.2a routines.
All images were corrected for bias, dark, and flat-field. 
The photometry is based on point-spread-function-fitting (PSF). For each image,
we selected at least 20 isolated stars to calculate the individual PSFs. 
A Moffat15 function fitted the observations best. In the following, only stars 
were used which are detected on all frames.
Because of instrumentally induced offsets and different air masses between 
the single frames, photometric reduction of each frame was performed separately 
and the measurements were then averaged and weighted by their individual
photometric error. The used photometric errors are based on the photon noise and the 
goodness of the PSF fit as described in \citet{Stet87}.

For NGC~6205 three different, overlapping, fields around the centre were observed. No significant photometric
offsets between the fields were detected. The most inner parts (radius of about 1$\farcm$5) 
of NGC~104 and NGC~7099 were not used for the further analysis because of the severe crowding and the
unresolved single star content.

\begin{table*}
\caption{Summary of results. The errors in the final digits of the corresponding quantity are given in parenthesis.}
\label{results_gcl}
\begin{center}
\begin{tabular}{lccc} 
\hline\hline 
& NGC 104 & NGC 6205 & NGC 7099 \\
\hline
$V = a + b(y)$ & $-$3.23(21)/0.909(22) & $-$2.82(16)/0.965(9) & $-$5.10(19)/1.009(17) \\
Reference & \citet{Hess87} & \citet{Grun98} & \citet{Alca98} \\
$a_{\rm 0} = a + b(g_{\rm 1} - y)$ & 0.922(3)/0.194(4) & 0.373(1)/0.217(9) & 0.946(2)/0.248(3) \\
3$\sigma$ (mag) & 0.033 & 0.042 & 0.028 \\
$n$(obj) & 1107 & 365 & 794 \\
$n$(positive) & 21 & 28 &  12 \\
$n$(negative) & 16 & 5 & 8 \\
$n$(frames) & 26 & 37 & 46 \\
\hline 
\end{tabular}
\end{center}
\end{table*}

\begin{figure}
\begin{center}
\includegraphics[width=80mm]{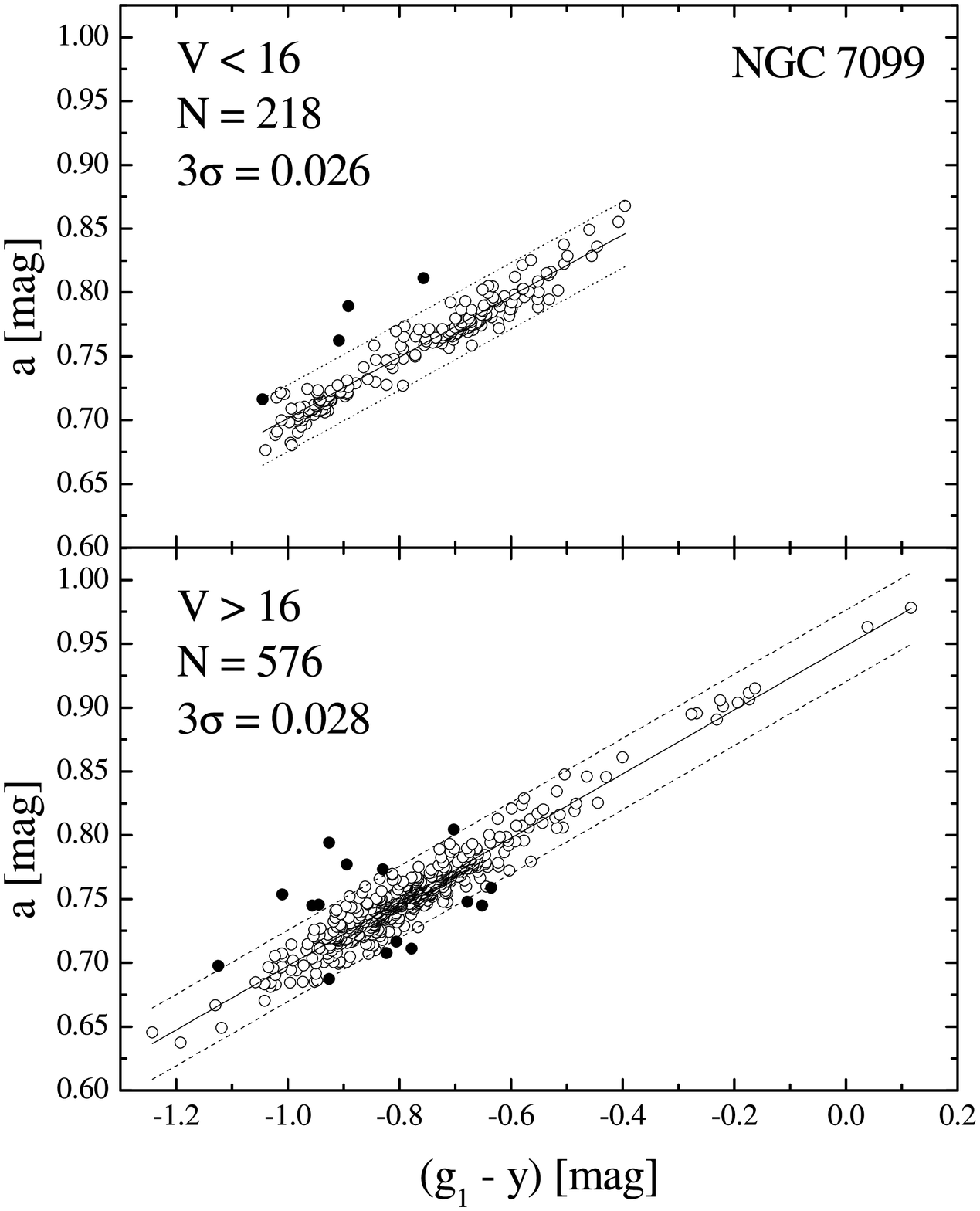}
\caption{The $a$ versus $(g_{\rm 1} - y)$ diagram of NGC 7099. The sample was divided into stars which 
are brighter (upper panel) and fainter (lower panel) than 16th magnitude.}
\label{ngc7099}
\end{center}
\end{figure}

The $a$-index is defined as
\begin{equation}
a = g_{\rm 2} - \frac{(g_{\rm 1} + y)}{2}.
\end{equation}

Since this quantity is slightly dependent on temperature
(increasing towards lower temperatures), the intrinsic peculiarity 
index ($\Delta a$) has to be defined as the difference between the
$a$-value -- individual measured $a$ for studied star -- and $a_{\rm 0}$ for non-peculiar star of the same colour. The locus
of the $a_{\rm 0}$-values is called normality line. Due to the reddening \citep{Mait93}, the 
$a$-values have to be corrected by 0.05$E(b-y)$. Because there
is no significant reddening towards our targets (see Table \ref{parameters_gcl}),
we neglect this effect for the further analysis. Assuming that all stars exhibit 
the same interstellar reddening and metallicity, peculiar objects deviate from the 
normality line more than 3$\sigma$.

For the transformation of the instrumental $y$ to $V$ magnitudes, we used the 
following references: \citet[NGC 104]{Hess87}, \citet[NGC 6205]{Grun98}, and
\citet[NGC 7099]{Alca98}. We note that the zero-points for the measurements 
taken at BNAO and CASLEO are not the same due to different CCD gain and bias levels
as well as extinction coefficients.

All results are summarized in Table \ref{results_gcl}. In total, 2266 stars on 109 frames
are finally analysed. The complete photometric data (Table \ref{photometry}) together with the coordinates
are available in electronic form.

\begin{table*}
\caption{The complete photometric data of the programme stars$^{1}$.}
\label{photometry}
\begin{center}
\begin{tabular}{rccccccccc} 
\hline\hline 
Id. & $\alpha$(2000) & $\delta$(2000) & $(g_{\rm 1} - y)$ & $\sigma(g_{\rm 1} - y)$ & $a$   & $\sigma a$ & $\Delta a$ & $V$   & $\sigma V$\\
    &                &                & (mag)             & (mag)                   & (mag) & (mag)      & (mag)      & (mag) & (mag) \\
\hline
NGC~104 \\
1	&	00:24:50.00	&	-72:05:15.2	&	-0.886	&	0.010	&	0.767	&	0.008	&	+0.017  	&	15.047	&	0.005	\\
2	&	00:24:49.69	&	-72:05:28.6	&	-0.904	&	0.004	&	0.768	&	0.003	&	+0.021  	&	13.872	&	0.002	\\
3	&	00:24:49.48	&	-72:05:46.4	&	-0.910	&	0.002	&	0.759	&	0.002	&	+0.013  	&	14.049	&	0.002	\\
4	&	00:24:49.41	&	-72:05:54.5	&	-0.838	&	0.001	&	0.762	&	0.002	&	+0.002  	&	13.572	&	0.001	\\
5	&	00:24:49.20	&	-72:05:38.5	&	-0.959	&	0.002	&	0.743	&	0.001	&	+0.008	  &	14.722	&	0.002	\\
6	&	00:24:48.58	&	-72:05:58.5	&	-0.869	&	0.001	&	0.745	&	0.002	&	-0.009   	&	13.621	&	0.001	\\
7	&	00:24:47.94	&	-72:04:55.5	&	-0.950	&	0.002	&	0.746	&	0.003	&	+0.008	  &	14.851	&	0.002	\\
\dots & \dots & \dots & \dots & \dots & \dots & \dots & \dots & \dots & \dots \\
\hline 
\multicolumn{9}{l}{$^{1}$A portion of the table is shown here for guidance. The complete table will be available online.}
\end{tabular}
\end{center}
\end{table*}

\section{Discussion}

The tool of $\Delta a$ photometry measures any flux/spectral abnormalities in the 5200\AA\,region. 
Employing the $\Delta a$ photometric system on globular clusters aims primarily 
towards two widely different star groups:
\begin{enumerate}
	\item Photospheric CP HB stars as found by \citet{Behr99}
	\item Peculiar RGB-, G- and K-type MS stars \citep{Grat12}
\end{enumerate}
The first group shows enhancements of iron peak elements of up to three times solar, or 2 dex compared to the mean metallicity, 
whereas the latter could be detected by peculiarities of Mg I lines as well as MgH features around 5200\AA. In addition, cool type
dwarfs and giants can be sorted out by the different equivalent widths of their Mg features. However, such a distinction can
also be easily done in the classical colour-magnitude diagram.

In Fig. \ref{cmds}, we present the results of our photometric observations. For each globular cluster, the 
$V$ versus $(g_{\rm 1} - y)$ and $a$ versus $(g_{\rm 1} - y)$ diagrams are shown. As expected, the 
colour-magnitude diagrams, especially the characteristics of the HB, of the three aggregates are widely 
different.

The slopes of the normality lines range from 0.194 to 0.248 (Table \ref{results_gcl}), which is perfectly
in line with values found for Galactic open clusters \citep{Neto07}. We notice that there seems to be a
correlation with [Fe/H], i.e. NGC 104 with the highest metallicity exhibits the shallowest slope. However,
with only three aggregates, one has to be careful with such a conclusion. Further observations are
clearly needed to prove this apparent correlation.

Due to the photon noise, fainter stars have, in general, larger photometric errors. Inspecting the
$a$ versus $(g_{\rm 1} - y)$ diagrams in Fig. \ref{cmds}, no correlation of the 3$\sigma$ detection limit
with the $a$ values is visible. For NGC 7099, the detailed analysis is shown in Fig. \ref{ngc7099}.
The complete sample was divided in stars brighter and fainter than 16th magnitude. Both samples are rather different in terms of
the number of stars and the range of $(g_{\rm 1} - y)$. In comparison with the
overall solution, only one object would not be detected as peculiar in the second sample with 
$\Delta a$\,=\,+0.027\,mag which is just one mmag below the detection limit. 
In addition, we divided the other samples in different $V$ subsamples and calculated the
3$\sigma$ detection limit anew. All values agree within 1.5\,mmag. Those tests justify using the sample
as a whole for our analysis. This strong advantage of the $\Delta a$ photometric system 
was already noticed before and is because for a given $(g_{\rm 1} - y)$ value, a wide range of $V$ values are sampled.

\begin{figure}
\begin{center}
\includegraphics[width=80mm]{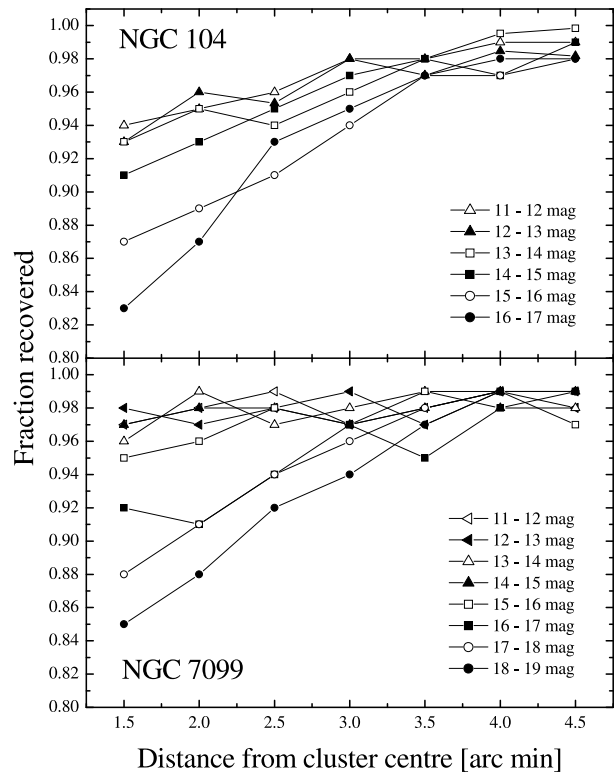}
\caption{Results of the artificial-star tests for NGC~104 and NGC~7099. The inner most core parts with a 
radius of about 1$\farcm$5 were not used for our analysis.}
\label{ats}
\end{center}
\end{figure}

\begin{figure}
\begin{center}
\includegraphics[width=80mm]{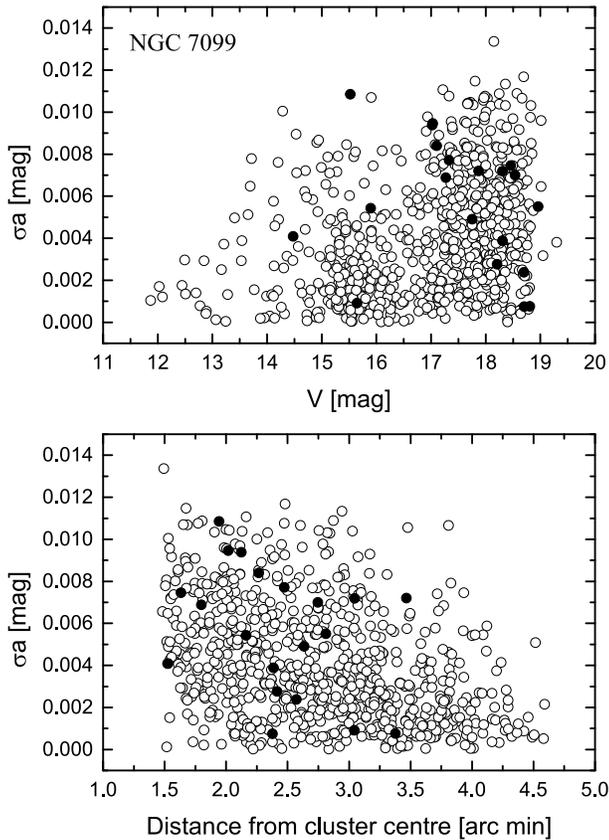}
\caption{The differences of the $a$ values from the artificial-star test and the observed ones ($\sigma a$) versus
the $V$ magnitude (upper panel) and the distance from the centre (lower panel) for NGC~7099. The peculiar candidates 
(Fig. \ref{cmds}) are marked as filled circles.}
\label{diff_ats}
\end{center}
\end{figure}

\begin{figure}
\begin{center}
\includegraphics[width=80mm]{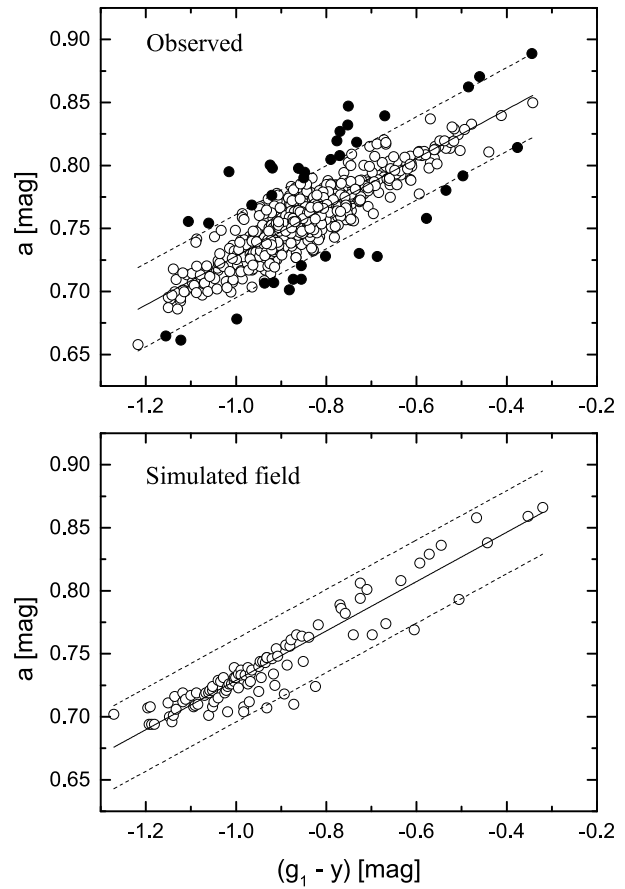}
\caption{The simulated fore- and background population using the theoretical Galactic model, TRILEGAL (lower panel)
and the observed one for NGC~104.}
\label{trilegal}
\end{center}
\end{figure}

We also performed an artificial-star test to determine the completeness level of
our sample which is very important when analysing crowded field photometry \citep{Ande08}. Normally, faint stars in very crowded 
regions are either lost in the saturated cores or have be detected against the high background of these bright star aureoles.
Thus, the magnitude limits for the detection of faint stars and the undercount correction estimates are functions of both the
stellar magnitude and the distance of the objects from the most crowded and therefore bright cluster centre.
As mentioned before, the inner most core parts with a radius of about 1$\farcm$5 were not used for our analysis. 
For this purpose, we added artificial stars with the IRAF task `ADDSTAR' to frames of each filter with the longest
integration times. About 100 independent experiments, each consisting of 1000 artificial stars within a 1\,mag
interval randomly scattered throughout the image, were performed. These frames were then photometrically analysed the
same way as the original ones. The detected fraction of artificial stars was determined in concentric annuli of 0$\farcm$5 width. 
Finally, a weighted average of the recovery fraction at each radius and magnitude interval was computed. For NGC~6205,
we got almost a 99\% completeness level for all bins because we observed fields quite far off the centre. Fig. \ref{ats}
shows the artificial-star tests for NGC~104 and NGC~7099. The band width of the individual curves is about $\pm$3\%.
Since almost all regions and magnitude ranges are well above 90\%, we are confident that the effect of undetected binary
stars do not play a significant role for our analysis. As a further test, the differences of the $a$ values from the 
artificial-star test and the observed ones were calculated. Fig. \ref{diff_ats} shows these differences versus the $V$ 
magnitude and the distance from the centre for the data of NGC~7099. Again, the distribution of the outliers does not 
significantly alter from the apparent normal type objects.

In total, photometric $\Delta a$ values for 2266 stars were secured. According to the well established
3$\sigma$ detection limit, 61 stars fall above, and 29 objects below the normality line. 
The latter can be also due to higher reddened background stars, which would 
shift these objects below the normality line. The only reliable distinction could be done via membership
probabilities based on proper motions. However, such kinematic data are not available for our target sample. Indeed, such field 
stars are present in globular cluster areas \citep{Sari12}. Those field stars are clearly visible in the right most
upper panel of Fig. 5 in \citet{Sari12} where they exhibit redder $(V-I)$ colours than the cluster members. In the $\Delta a$ photometric diagram,
such stars could lie below the normality line, but not above it. We also checked the possibility 
of the influence of undetected visual companions for the outliers. In general, applying PSF-fitting, already
accounts for such cases. A comparison of `visual binaries' among normal type objects and outliers shows no
significant accumulation among the latter. 

To estimate the $a$ versus $(g_{\rm 1} - y)$ diagram of background/foreground stars
in the field of view of our targets, we used the theoretical Galactic model, TRILEGAL 1.6\footnote{http://stev.oapd.inaf.it/trilegal} 
described by \citet{Gira05}. It includes the populations of the thin and thick disc as well as the Galactic halo but it is not able
to simulate star clusters, yet. We have simulated fields of 12'x12' with the central coordinates as listed in Table \ref{parameters_gcl}.
Up to now, the $\Delta a$ photometric system is not included in the list of systems. As workaround, we used the synthetic $\Delta a$ photometry 
taken from the Vienna New Model Grid of Stellar Atmospheres, NEMO\footnote{http://www.univie.ac.at/nemo} \citep{Heit02}. The TRILEGAL
output includes \Teff, \logg, [Fe/H], and $V$ for each object. First of all, we restricted the sample to the $V$ magnitudes as deduced
from Fig. \ref{cmds}. Then, for each star, we searched for the closest model from the NEMO data base and took the corresponding 
$\Delta a$ photometry. Fig. \ref{trilegal} shows the observed and synthesized field of NGC~104. The situation for the other two fields
is similar. The slope for the normality line of the observational data is 0.194(4) whereas it is 0.196(5) for the synthetic data. There are
a few stars below (see discussion above) but none above the normality line. We are, therefore, confident that fore- and background stars
cannot mimic a statistically significant number of peculiar globular cluster members in the $a$ versus $(g_{\rm 1} - y)$ diagram. However, some
negative outliers of NGC~104 and NGC~7099 (Fig. \ref{cmds}) can be caused by field stars.

{\it NGC~104:} no significant deviating $\Delta a$ values for HB stars were found. About 10 outstanding objects
are probable non-members and can be easily identified in the $V$ versus $(g_{\rm 1} - y)$ diagram.

{\it NGC~6205:} this is the most detailed investigated globular cluster among our targets. We measured three HB stars
listed in \citet{Behr03}, namely, WF2-820 (No. 147), WF-2692 (No. 227), and WF2-3035 (No. 74). The latter shows no
chemical peculiarities, whereas the other two have large overabundances of almost all iron peak elements from 1 to 1.5
dex compared to the cluster metallicity. Our $\Delta a$ values are perfectly in line with the abundances. For WF2-3035
we find an insignificant value of +8 mmag, whereas the other two stars were detected with +57 and +60 mmag. This lends to
confidence that the $\Delta a$ photometric system is indeed capable to detect CP HB stars. However,
further observations of such objects have to prove this conclusion. There are also several BHB stars which are
below the normality line. For none of these objects, membership probabilities are available in the literature \citep{John12}.
We may speculate that this behaviour could be due to photometric variations. Such a behaviour is a common phenomenon for 
CP stars \citep{Paun11b}, but was never investigated for members of
globular clusters, yet.

As next step, we investigated the RGB
stars published by \citet{John05} among our sample. In total, we find 10 stars in common. The [Mg/Fe] values for
those objects range from $-$0.15 to +0.30 dex. None of them exhibit a significant $\Delta a$ value, probably because 
the effect of the Mg lines compared to Fe for such rather low peculiarities is too small in the 5200\AA\, region.

{\it NGC~7099:} there are no detailed elemental abundances for members in the literature available. There are three
HB stars with a $\Delta a$ detection, from which one object lies significantly above the HB in the $V$ versus $(g_{\rm 1} - y)$ 
diagram. If it is a member then this is probably a very interesting object for follow-up observations. The reasons why several fainter
stars deviate from the normality line (as also seen for NGC~104), are not straightforward to determine without any additional
observations. However, from our previous considerations one can think of non-members or very strong peculiarities of magnesium.

\section{Conclusions and outlook}

We presented, for the first time, photometric $\Delta a$ observations of three globular clusters. 
It measures the flux distribution in the region from 4900 to 5600\AA\, \citep{Paun14}.
This three filter narrow band system was originally developed to detect classical CP
stars of the upper MS. Later on, it turned out that is also capable to detect underabundant,
emission as well as shell type stars. 

Another mile stone was its extension to field stars and clusters in the Large Magellanic Cloud. Even in this
underabundant (compared to the Milky Way) global environment, we were able to detect CP stars, which were later
confirmed spectroscopically \citep{Paun11a}.

In total, we present photometry of 2266 stars from 109 individual frames. According to the 3$\sigma$
detection limit of each globular cluster, we find 61 objects with positive and 29 with negative
$\Delta a$ values. This corresponds to an upper limit of about 3\% of apparent peculiar objects.

For NGC 6205, we were able to compare our results with abundance determinations from the literature.
The $\Delta a$ values of three HB stars, one without chemical peculiarities, listed in \citet{Behr03},
are in perfect agreement. The peculiar objects were clearly detected with +57 and +60 mmag, whereas
the non-peculiar object does not stand out. In addition, we analysed 10 RGB stars with [Mg/Fe] values 
from $-$0.15 to +0.30 dex, published by \citet{John05}. None of them exhibit a significant $\Delta a$ value, which
means that the elemental peculiarity of magnesium has to be much larger to be detected.

Several future steps have to be still performed. First of all, further $\Delta a$ observations
of globular clusters and members with known elemental peculiarities are needed to test and establish
the system and its results. The presented apparent peculiar objects should be analysed in detail, using 
high resolution spectroscopy and corresponding stellar atmospheres. However, $\Delta a$ seems to be a very efficient way to preselect 
such very interesting objects by means of photometry. As last step, current available stellar atmospheres within
the investigated astrophysical parameter space should be used to calculate synthetic colours for a comparison
with the observations.

\section*{Acknowledgements}
This project is financed by the SoMoPro II programme (3SGA5916). The research leading
to these results has acquired a financial grant from the People Programme
(Marie Curie action) of the Seventh Framework Programme of EU according to the REA Grant
Agreement No. 291782. The research is further co-financed by the South-Moravian Region. 
It was also supported by the grants GA \v{C}R P209/12/0217, 14-26115P, 7AMB14AT015, and
the financial contributions of the Austrian Agency for International 
Cooperation in Education and Research (BG-03/2013 and CZ-09/2014).
IKhI acknowledges partial support from NSF grants DO 02-85 and 01/7 - DNTS AT.
This work reflects only the author's views and the European 
Union is not liable for any use that may be made of the information contained therein.

\label{lastpage}
\end{document}